\documentclass[useAMS,usenatbib]{mn2e}

\usepackage[utf8]{inputenc}
\usepackage[T1]{fontenc}
\usepackage{aecompl}
\usepackage{url}
\usepackage[english]{babel}
\usepackage{array}
\usepackage{multirow}
\newcolumntype{C}[1]{>{\centering\let\newline\\\arraybackslash\hspace{0pt}}m{#1}}

\usepackage{natbib}
\usepackage{amsmath}
\usepackage{ifpdf} 
\ifpdf
  \usepackage{graphicx}
  \usepackage{epstopdf}
\else
  \usepackage{graphicx}
\fi
\newcommand{\fas}{\mbox{\ensuremath{.\!\!^{\prime\prime}}}}
\newcommand{\uJy}{\,\ensuremath{\umu}\text{Jy}{}}
\newcommand{\kvega}{\ensuremath{K_\mathrm{Vega}}}
\newcommand{\ksmp}{K20}
\newcommand{\msun}{\ensuremath{\mathrm{M}_\odot}}
\newcommand{\msunyr}{\ensuremath{\msun\,\mathrm{yr}^{-1}}}
\newcommand{\sfd}{\ensuremath{\Sigma_\mathrm{SFR}}}
\newcommand{\sfdu}{\ensuremath{\msun\,\mathrm{yr}^{-1}\,\mathrm{kpc}^{-2}}}
\newcommand{\lsun}{\ensuremath{\mathrm{L}_\odot}}
\newcommand{\lir}{\ensuremath{L_\mathrm{IR}}}

\title[Estimating sizes of faint, distant galaxies in the submillimetre regime]{Estimating sizes of faint, distant galaxies in the submillimetre regime}
\author[L. Lindroos et al.]
{\parbox{\textwidth}{L. Lindroos$^{1}$\thanks{E-mail: lindroos@chalmers.se},
K. K. Knudsen$^{1}$,
L. Fan$^{2}$,
J. Conway$^{1}$,
K. Coppin$^{3}$,
R. Decarli$^{4}$,
G. Drouart$^5$,
J. A. Hodge$^{6}$,
A. Karim$^{7}$,
J. M. Simpson$^{8}$,
and J. Wardlow$^{8,9}$}\vspace{0.4cm}
\\
\parbox{\textwidth}{
$^{1}$Department of Earth and Space Sciences, Chalmers University of Technology, Onsala Space Observatory, SE-439 92 Onsala, Sweden\\
$^{2}$Shandong Provincial Key Lab of Optical Astronomy and Solar-Terrestrial Environment, Institute of Space Science,\\\,\,\,Shandong University, Weihai, 264209, China\\
$^{3}$Centre for Astrophysics Research, University of Hertfordshire, College Lane, Hatfield, AL10 9AB, UK\\
$^{4}$Max-Planck-Institut f\"{u}r Astronomie, K\"{o}onigstuhl 17, D-69117 Heidelberg, Germany\\
$^{5}$International Centre for Radio Astronomy Research, Curtin University, GPO Box U1987, Perth, WA 6845, Australia\\
$^{6}$Leiden Observatory, Leiden University, P.O. Box 9513, 2300 RA Leiden, The Netherlands\\
$^{7}$Argelander-Institut für Astronomie, Universität Bonn, Auf dem Hügel 71, D-53121 Bonn, Germany\\
$^{8}$Centre for Extragalactic Astronomy, Department of Physics, Durham University, South Road, Durham DH1 3LE, UK\\
$^{9}$Dark Cosmology Centre, Niels Bohr Institute, University of Copenhagen, DK-2100 Copenhagen, Denmark }}

\begin{document}

\date{Accepted 2016 July 05. Received 2016 June 25; in original form 2016 December 21}

\pagerange{\pageref{firstpage}--\pageref{lastpage}} \pubyear{2016}

\maketitle

\label{firstpage}

\begin{abstract}
	We measure the sizes of redshift $\sim2$ star-forming galaxies by stacking data from the Atacama Large Millimeter/submillimeter Array (ALMA).
	We use a $uv$-stacking algorithm in combination with model fitting in the $uv$-domain
	and show that this allows for robust measures of the sizes of marginally resolved sources.
	The analysis is primarily based on the 344 GHz ALMA continuum observations centred on 88 sub-millimeter galaxies in the LABOCA ECDFS Submillimeter Survey (ALESS).
	We study several samples of galaxies at $z\approx2$ with $M_*\approx5\times10^{10}\msun$,
	selected using near-infrared photometry (distant red galaxies, extremely red objects, sBzK-galaxies, and galaxies selected on photometric redshift).
	We find that the typical sizes of these galaxies are $\sim0\fas6$ which corresponds to $\sim5$ kpc at $z=2$,
	this agrees well with the median sizes measured in the near-infrared $z$-band ($\sim0\fas6$).

	We find errors on our size estimates of $\sim0\fas1-0\fas2$,
	which agree well with the expected errors for model fitting at the given signal-to-noise ratio.
	With the $uv$-coverage of our observations (18-160 m),
	the size and flux density measurements are sensitive to scales out to 2\arcsec.
	We compare this to a simulated ALMA Cycle 3 dataset with intermediate length baseline coverage,
	and we find that, using only these baselines, the measured  stacked flux density would be an order of magnitude fainter.
	This highlights the importance of short baselines to recover the full flux density of high-redshift galaxies.
\end{abstract}


\section{INTRODUCTION}
The star-formation rate density in the universe peaks at $z\sim2$ \citep[e.g.][]{madau2014},
making this a very important epoch in the formation of galaxies.
For galaxies at these redshifts submillimeter (sub-mm) emission is a commonly used tracer of star formation
\citep[e.g.][]{daddi2010b},
often used in combination with ultraviolet and optical measurements to allow reliable star-formation rate (SFR)
estimates for galaxies with very different dust properties \citep[e.g.][]{tacconi2013,dacunha2015}.
The Atacama Large millimeter/submillimeter Array (ALMA) 
and IRAM NOrthern Extended Millimeter Array (NOEMA)
are currently producing a large wealth of data at frequencies of $200-350$ GHz,
allowing us to measure the sub-mm emission from high-redshift galaxies previously to faint to study.
Observing at these frequencies is efficient for high redshifts,
as the flux density for galaxies at a given SFR is expected to be almost constant for redshift z $\sim1-6$
due to the negative $K$-correction \citep[e.g.][]{blain2002, casey2014}.

Current observations with ALMA and NOEMA primarily focus on the galaxies with high SFR, $>100$ \msunyr{},
however, these galaxies constitute a small fraction of the total star formation \citep[e.g.][]{bouwens2011, rodighiero2011}.
It is possible to study single sources from much fainter galaxy populations,
e.g., with 50 ALMA antennas and $\sim1$ hour integration we can reach a depth of 20\uJy{}/beam  at 345 GHz,
which corresponds to 1 $\sigma$ uncertainty of $\sim 2 \msunyr$ at $z=2$.
However, to obtain large samples of galaxies for statistical studies is very expensive.
An alternate approach is to study galaxies that are amplified by gravitational lensing.
By using lensing it is possible to detect very faint sources with shorter observations, e.g., 
\cite{watson2015} detected a $z\sim7$ galaxy with a SFR of 9 $\msunyr$ and a flux density of 0.61 mJy at 220 GHz,
which would require only a $\sim$30 s integration for a 5$\sigma$ detection with 50 ALMA antennas.
However, it can be difficult to obtain large samples of such galaxies as high magnifications are rare.
A third approach is stacking,
which uses shallower surveys to study statistical properties of large samples galaxies which have previously been detected at other wavelengths.
Stacking is a common technique used across many different wavelength:
$\gamma$-rays \citep[e.g.][]{aleksic2011},
X-rays \citep{chaudhary2012, george2012},
optical/near infrared \citep{zibetti2007,matsuda2012,gonzalez2012},
mid/far infrared \citep[e.g.][]{dole2006},
and radio \citep{boyle2007,ivison2007,hodge2008,hodge2009,dunne2009,karim2011}.

Looking specifically at sub-mm emission,
stacking has been applied to data from James Clerk Maxwell Telescope (JCMT) and Atacama Pathfinder EXperiment (APEX),
using several different samples of high-redshift galaxies,
\citep[e.g.][]{webb2004,knudsen2005,greve2010}.
Compared to these surveys,
ALMA can achieve sub-arcsecond resolution,
which is orders of magnitude better than the $19\fas2$ and $15\fas0$ at $345\,$GHz of APEX and JCMT respectively.
Firstly,
this allows us to measure the flux density of the sources without being affected by confusion,
which is believed to impact the result of stacking at JCMT and APEX resolutions \citep[e.g.][]{webb2004}.
Secondly,
we can study the structure of our stacked source.
Several studies have found star-forming galaxies at redshifts of $z\sim2$ have large sizes,
e.g. \cite{daddi2010b} found sizes up to 1\fas5 for sample of $z\sim2$ galaxies.

\cite{decarli2014} used stacking to measure the sub-mm flux density of star-forming galaxies in the Extended Chandra Deep Field South (ECDFS)
with data from the ALMA.
In this paper we will build on the work by \cite{decarli2014},
using the same data,
but extending the analysis to focus on the sizes of the stacked sources.
\cite{decarli2014} performed stacking on the imaged pointings,
analogous to how stacking is done at other wavelengths.
However, as seen in \cite{lindroos2015}, this may not be ideal for interferometric data.
In this paper we instead adopt the $uv$-stacking approach described in \cite{lindroos2015},
which performs the stacking directly on the visibility data.
When using image stacking in mosaiced data sets,
it is necessary to combine data from pointings imaged with different restoring beams.
Because of this,
it is very difficult to deconvolve the source structure from the beam in the final stacked image.
Using $uv$-stacking, we combine the data in the $uv$-domain,
and the beam can be directly calculated from the new $uv$-coverage.
Therefore,
using the $uv$-stacking algorithm is especially important for measuring the sizes of the stacked sources.

While the work in this paper is primarily focused on stacking high-redshift galaxies,
the stacking techniques applied are quite general.
Many of the lessons learned apply to any ALMA stacking of marginally extended sources.

The paper is structured as following.
In \S 2, we describe the ALMA data we use and in \S 3 we describe the sample,
as well as the photometric near infrared and optical catalogue.
In \S 4 we describe a set of simulations performed to test various aspect of the stacking result
and in \S 5, we describe our $uv$-stacking procedure.
In \S 6, we summarise our results, including the typical galaxy sizes for each sample.
Finally, in \S 7 we discuss the implications of the results both for star formation at $z\sim2$,
and for general stacking of ALMA data.

In this paper we use a standard cosmology with $H_0 = 67.3\,\mathrm{km}\,\mathrm{s}^{-1}\,\mathrm{Mpc}^{-1}$,
$\Omega_\Lambda = 0.685$, and $\Omega_m = 0.315$ \citep{planck2013}.
All magnitudes are in AB \citep{oke1974} unless otherwise specified.

%

\section{DATA}
Our analysis is based on data from the ALMA survey of the submillimetre galaxies (SMGs) detected in LESS (ALESS, \citealt{hodge2013}),
where LESS is the LABOCA ECDFS Submm Survey,
LABOCA is the Large Apex BOlometer CAmera mounted on APEX,
and ECDFS is the Extended Chandra Deep Field South.
The ALESS survey is composed of 122 pointings across the ECDFS,
centred on 122 SMGs,
observed during ALMA Cycle 0 between October and November 2011.
The observations are tuned to a frequency of 344 GHz and have a typical resolution around 1\fas6 $\times$ 1\fas2.
The median value of the noise (standard deviation) in the centre of each pointing is $\sim0.4$mJy/beam.
All pointings with central noise $>0.6$ mJy/beam or beam axis ratio $>2$ are excluded from the analysis,
see \cite{hodge2013} for more details.
As such our data consist of 88 ``good quality'' pointings,
each with a field of view (full width at half power of the ALMA primary beam)
of 17\fas3 at 344 GHz,
covering a total of 5.6 arcmin$^2$.

\section{PHOTOMETRIC GALAXY SELECTION}
In this paper we extend the analysis of \cite{decarli2014},
using the same sample selection.
The selection is based on the photometric catalogue of the ECDFS assembled using the same procedure as \cite{simpson2014},
using primarily data from the Wide MUlti-wavelength Survey by Yale-Chile
(MUSYC; \citealt{taylor2009a}).
The MUSYC catalogue is a $K$-band flux limited sample,
covering a $30\arcmin\times{}30\arcmin$ area of the ECDFS,
with photometry for the sources in the bands $UBVRIzJHK$.
At $K_\mathrm{AB} = 22$ mag the sample is 100 per cent complete for point sources,
and 96 per cent complete for extended sources with a scale radius of 0\fas5.
\cite{simpson2014} extend the catalogue by including a deep $J$ and $K$ band catalogue Zibetti et. al (in preparation),
the Taiwan ECDFS NIR survey
\citep{hsieh2012}, and Spitzer/IRAC 3.6, 4.5, 5.8, and 8.0 \micron{} 
images from the Spitzer IRAC/MUSYC Public Legacy Survey
\citep{damen2011}.
From the MUSYC data we also have estimates of the photometric redshifts for our galaxies,
estimated using EAZY \citep{brammer2008}.

From the catalogue we select four different samples.
We limit all samples with $\kvega < 20$, $z>1$, and further limit the samples as follows:

\begin{enumerate}
	\item All sources with $(z-K -0.04) > 0.3 (B-z+0.56)-0.5$ which separate the galaxies from the stars \citep{daddi2004}. This sample was refered to the $\kvega < 20$ sample in \cite{decarli2014} and will be refered to as the \ksmp{} sample in this paper.
	\item Actively star-forming galaxies selected using the sBzK criteria by \cite{daddi2004}, i.e., $(z-K-0.04) - (B-z+0.56) > -0.2$. 
	\item Distant Red Galaxies (DRGs) selected using $J-K > 1.32$ \citep{franx2003}.
	\item Extremely Red Objects (EROs) selected using $(R-K)>3.35$ and $(J-K) > 0.1$ \citep{elston1988}.
\end{enumerate}

This results in our samples being the same as the $z>1$ samples in \cite{decarli2014}.

\begin{figure}
	\vbox to90mm{\vfil \includegraphics[width=85mm]{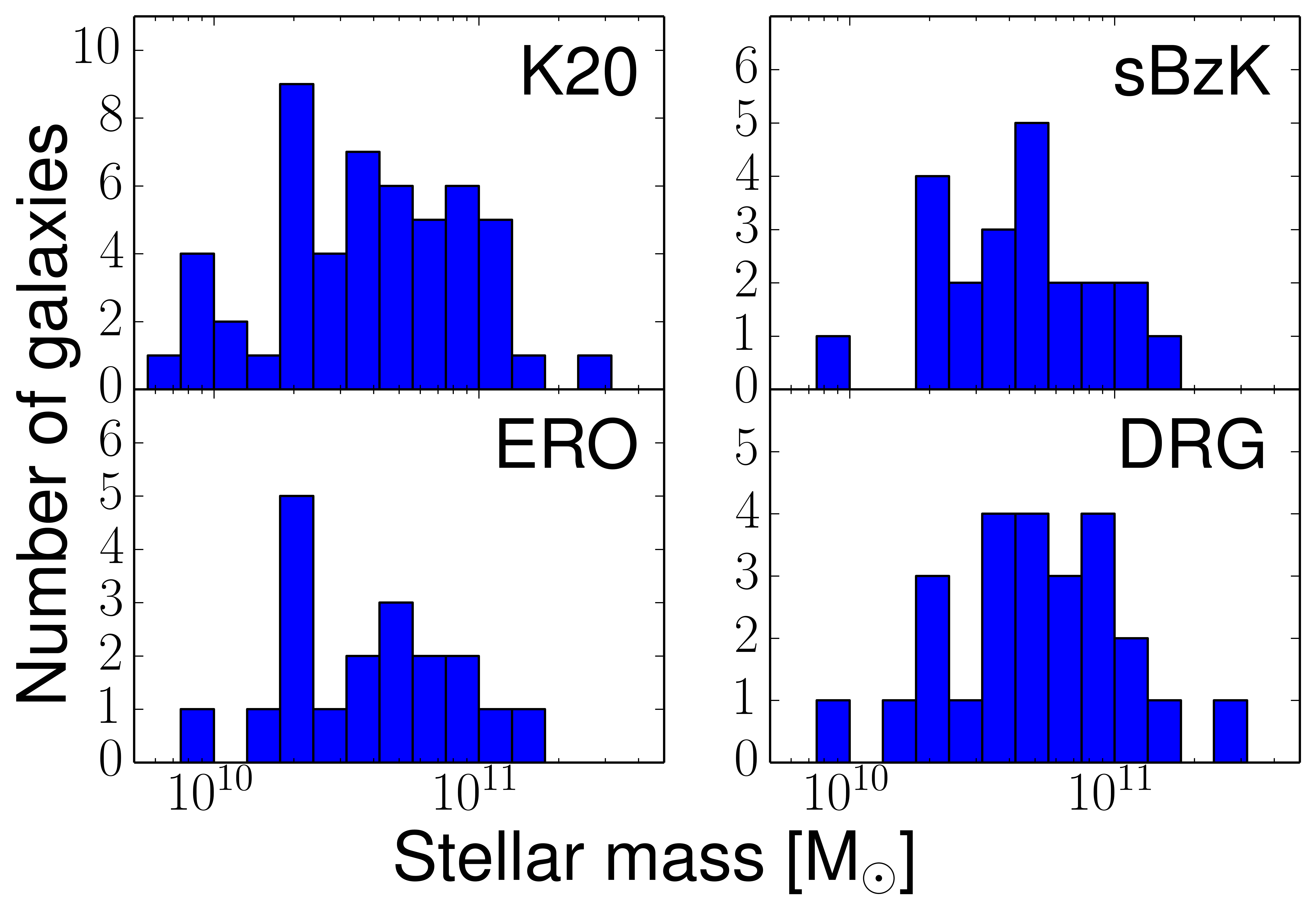}
	\caption{Distribution of stellar masses for each sample. 
		The stellar masses are estimated by SED fitting to optical and near-infrared band using PEGASE 2 \citep{fioc1997}.
	\label{fig:stellar_mass} }
	\vfil}
\end{figure}

Using the MUSYC photometry we also estimate the stellar mass ($M_*$) of our selected galaxies.
The stellar-mass estimates are done using PEGASE 2 \citep{fioc1997}.
For each galaxy we use all available bands, i.e., $U$, $B$, $V$, $R$, $I$, $z$, $J$, $H$, and $K$.
Using four different galaxy templates (elliptical, spiral Sa, spiral Sd, and starburst),
all assuming Kroupa IMF,
we fit for stellar mass. 
The redshift is not fitted directly, instead we use the photometric estimates from \cite{taylor2009a}.
For each source we choose the model with the lowest $\chi^2$,
with more than 90 per cent of sources best fitted by the elliptical or the starburst model.
The distributions of stellar masses for these samples are shown in Fig. \ref{fig:stellar_mass}.

\section{STACKING ROUTINE}
\label{sec:method}
The samples are stacked using the $uv$-stacking algorithm described in \cite{lindroos2015}.
The algorithm performs the stacking operation directly on the visibility data.
We use model fitting in $uv$-domain to estimate the flux densities and sizes of our stacked sources.
For comparison with previous image stacking results we also use a simpler flux density estimate which assumes a point source,
where the flux density is estimated using the weighted average of all non-flagged visiblities.
We refer to this estimate at the \emph{point-source estimate}.

\subsection{Removing bright sources}
Prior to stacking each sample,
all bright sources not part of the sample are subtracted from the visibility data.

The modelling and subtraction was performed as follows.
The data is imaged and cleaned using Common Astronomy Software Applications package\footnote{http://casa.nrao.edu} (CASA) version 4.4.
Each pointing is imaged separately with a cell size of 0\fas2{} and 
cleaned down to a depth 1.8 mJy/beam.
A model for the bright sources is built from the clean model,
excluding all sources within 1\arcsec{} of a stacking position.
The model is subtracted from the $uv$-data,
to produce a residual data set.
To ensure that the visibility weights are accurate after the subtraction,
they weights are recalculated using the scatter of the visibilities in each baseline and time bin.

Note that the aim of the bright source subtraction is to remove bright sources that are unrelated 
to the target stacking sources,
not to remove those bright in the target sample.
As such, this subtraction is performed separately for each sample.
We also note that the bright source subtraction is based on the clean models,
which while not fully removing the sources,
is found to be sufficient for stacking,
see section \ref{sec:robust}.

\subsection{Weights and primary-beam effects}
The $uv$-stacking method prescribed in \cite{lindroos2015} uses a weighted average.
We calculate the stacking weights for each position from the primary beam attenuation.
Noise variations between pointings are included in the visibility weights,
and are thus not included in the stacking weights.
To ensure that the visibility weights are accurate,
they are recalculated prior to stacking from the scatter of each baseline and integration,

The primary beam attenuation ($A_N$) is estimated using the ALMA model present in CASA version 4.4, 
i.e., an Airy pattern with a full width at half maximum (FWHM) of 1.17$\frac{\lambda}{D} \approx 17\fas{}5$.
This results in stacking weights calculated as
\begin{equation}
	W_k = \left[A_N(\bmath{\hat{S}_k})\right]^2,
\end{equation}
where $W_k$ and $\bmath{\hat{S}_k}$ are the weight and position of source $k$.

\subsection{Model fitting}
\label{sec:model_fit}
Two different models are used to characterise the stacked sources.

The first: a point source model defined by
\begin{equation}
	V_\mathrm{ps}(u,v) = \Phi e^{2\pi i \frac{u l + v m}{\lambda}}
	\label{eq:model_ps}
\end{equation}
where 
$(u,v)$ are the projected baselines,
$\lambda$ is the wavelength,
$(l,m)$ are the direction cosines relative to the phase centre,
and $\Phi$ is the flux density of the source.

The second: a Gaussian model defined by
\begin{equation}
	V(u,v) = e^{\left(\pi^2 \frac{\Theta^2}{8 \ln 2} \times \frac{(u+v)^2}{\lambda^2}\right)} V_\mathrm{ps}(u,v)
\end{equation}
where 
$V_\mathrm{ps}$ is defined according to Equation \ref{eq:model_ps},
and $\Theta$ is the source size (FWHM) in radians.

The models are fitted in the $uv$-domain to our stacked sources using the least square minimizer package Ceres .
\footnote{Ceres \citep{ceres-solver}  uses a Levenberg-Marquardt algorithm \citep{levenberg1944} for non-linear least square minimization.
It supports several different solvers for the linear step.
We use the solver based on Cholesky decomposition,
which for our data set typically run 2 times faster compared to a standard QR factorisation.
The fit is terminated at the first to occur within 50 iterations,
a parameter change in the last step of less than $10^{-15}$,
or a relative $\chi^2$ change less than $10^{-18}$.
All flux densities are constrained to be positive.}
The model fitting is done to all non-flagged visibilities,
and includes the visibility weights in the $\chi^2$ minimization.

\subsection{Estimates of uncertainties in stacking}
\label{sec:uncertainties}
We use two different methods to estimate the uncertainties of our size and flux density estimates:
a Monte Carlo method where random sources are inserted into the data and stacked,
and a bootstrapping method.

The Monte Carlo simulation for a given sample and model is performed as follows:
a set of Monte Carlo sources is generated with the same number of sources as the given sample.
The position for each source is randomized,
however,
always within the same pointing as their corresponding actual source.
Each source is modelled as the fit for the given model to the stacked sources of the given sample.
The set of Monte Carlo sources are introduced into the residual data set for the given sample
and stacked using the same procedure as for the actual samples.
Finally the flux density and size of the stacked Monte Carlo sources are estimated using the given model.
This procedure is repeated a 100 times for each sample and model
to produce a distribution of estimated Monte Carlo flux densities and sizes.
The uncertainties are calculated as the standard deviation of our Monte Carlo estimates.

The bootstrapping method is performed by resampling the galaxies in each sample allowing replacements,
e.g., picking galaxy 1 two times and galaxy 2 one time, and galaxy 4 one time from a sample of 4 galaxies.
We stack the new sample, and estimate the flux density and size using model fitting.
By studying the distribution of the parameters in different resamples we can measure the influence of noise and underlying sample variance on the result.
To fully exhaust all possible resamplings would require $\binom{N\times{}2-1}{N}$ resamplings where N is the number of galaxies in the sample.
This is approximately $10^{12}$ for the sBzK sample,
however, we can get a good estimate using a much smaller number of resamplings.
As such we resample 1000 times for each target sample.
The error on each paramater is reported as where the measured cummulative distribution function (CDF) crosses 0.159 and 0.841,
equivalent to $\pm 1\sigma$ of a Normal distribution.
The estimated parameters are also recentered on where the measured CDF crosses 0.5,
thereby reducing the influence of outliers on the result.

We choose to refer to the first method as the Monte Carlo method as this is the same as the Monte Carlo method used in \cite{decarli2014}.
However, it is worth noting that the bootstrap method is also a Monte Carlo method as we do not fully
exhaust all possible resamples, however, in this work we will refer to it as bootstrapping.

\subsection{Estimate of uncertainties in model fitting}
\label{sec:visbootstrap}
The bootstrapping described in \S\ref{sec:uncertainties} uses resampling of the galaxies to estimate the uncertainties of stacking.
Using bootstrapping we can also estimate the uncertainty of the model fitting,
by resampling the visibilities of the $uv$-data.
This method is not used for the stacked results as it will not estimate uncertainty from variance within the sample,
however, it is powerful for model-fitting of individual sources.
We will refer to this method as visibility bootstrapping.

\section{SIMULATIONS AND MORPHOLOGY OF THE SUB-MM EMITTING REGION}
\label{sec:simulation}
The model fitting described in section \ref{sec:model_fit} 
allows us to estimate the total flux densities and typical sizes of our stacked sources.
The $uv$-models used aim to simulate the behaviour of the averages of our samples.
They are not based on the underlying morphologies of our samples.
However, 
looking at the data in the $uv$-domain we can obtain hints on the underlying structures of our sources.
We have simulated several possible morphologies for the galaxies of our samples,
to test if they produce different signatures in the stacked data,
and to be able to compare them with our actual stacked data.

For each simulation we generate a model of fake sources
and then simulate an ALMA data set with the following procedure.
We take the raw ALESS data set and set all visibilities to zero,
then we add the model and noise to the data set.
The noise is added using the simulator tool ({\tt sm}) in CASA,
using the default parameters which produces a realistic noise for the ALMA site.
After this the visibility weights are recalculated
by using the scatter in each baseline and time bin.

This simulated data set is then stacked using the same procedure as for our real data sets
(section \ref{sec:method}.)

\subsection{Clumpy morphology}
\label{sec:sim_clumps}
Observations of high-z star-forming galaxies at rest-frame wavelengths of $\sim$ 200 nm
indicate that they are more clumpy compared to their counterparts at lower redshifts
\citep[e.g.][]{im1999,schreiber2009}.
Based on this we have generated a model where all the sub-mm flux is coming from a few clumps.

For each source in the sample we generate 3 clumps, i.e., 3 point sources.
The clumps are scattered uniformly around the source position,
with a maximal distance of 0\fas6 from the centre.
Each clump is given a flux of 0.7 mJy and a size of $500\,$pc,
resulting in a total flux of 2.1 mJy for each simulated galaxy.

We simulate two different $uv$-coverages.
Firstly, the same as our ALESS observations,
with a similar level of noise added using the standard {\tt sm} parameters.
Secondly, an intermediate length baseline array with 36 antennas taken from ALMA Cycle 3:
the C36-5 configuration described in the ALMA Cycle 3 technical handbook \footnote{https://almascience.eso.org/documents-and-tools/cycle3/alma-technical-handbook},
with baselines from 45 m to 1.4 km.
The total observation time is scaled down to achieve a similar noise,
i.e. 1 h spread evenly over the 122 pointings.

\subsection{Following stellar morphology}
\label{sec:hstsim}
The inner parts of the ECDFS are covered by the GOODS-S survey \citep{Giavalisco2004},
with Hubble Space Telescope (\textit{HST}) observations in $z$-band (900nm) with a point-source sensitivity of 27.4 mag.
The wider field of ECDFS is observed in the Galaxy Evolution from Morphology and SEDs \citep[GEMS,][]{rix2004},
with \emph{HST} data in the F606W and F850LP filters, however, 
at a shallower depth compared to the GOODS-S observation: 2000 s typical integration as compared to 6000 s.
At $z\sim{}2$ the $z$-band observed corresponds to a rest-frame wavelength of approximately $300\,$nm,
where the emission is dominated by light from intermediate mass stars \citep{bruzual2003}.

In contrast the sub-mm emission observed by ALMA at 344 GHz will primarily trace star-formation surface density \citep{leroy2012}.
We can use this to test whether the star formation follows a significantly different morphology compared to the stellar population.
Since we are working with stacking we can not study individual galaxies,
however, we can say something about average properties.
As such, we produce a simulated dataset
where the star formation has the same surface density as the stellar mass traced by the HST $z$-band.
This simulated dataset can be directly compared with the actual stacked data.

The simulated dataset is produced as follows:
we select all sources which are part of the \ksmp{}
and have at least a 5$\sigma$ detection in either GEMS (band F850LP) or GOODS-S,
a total of 32 sources.
These galaxies are stacked using the same method as for the other samples.
For each source we take the HST image,
mask all pixels below 5 times the noise,
and scale to the same total flux density as the stacked average for the sample,
i.e., 1.4 mJy.
These images are then used as input model for a simulation,
following the same method as described for the clumpy model.

\section{STACKING RESULTS}
\subsection{Astrometry}
As part of the stacking process,
we re-align the astrometry from our optical catalogue with our ALMA astrometry.
From model fitting with a point source we find an offset in declination of approximately 0\fas3,
with small variations ($<0\fas1$) between different samples.
We also fit the position using the disk and Gaussian models,
finding a variation of $\sim$ 0\fas02 between the different models.
This is consistent with the offset found by \cite{simpson2014} for the bright galaxies in the same data.
Based on this, all stacked datasets were phase rotated with 0\fas3 in declination.

We also study random offset for individual positions.
Such random offsets can result from mis-registration of the positions in the optical catalogue
due to the limited signal-to-noise ratio (SNR) of the $K$-band detections.
Of the 100 sources in the \ksmp{} 11 galaxies are detected at a peak SNR $> 5$.
For these galaxies we estimate the peak position using a point-source model,
and the errors of the fitted positions using visibility bootstrapping (see \S\ref{sec:visbootstrap}).
We find that weighted means of offset between the optical positions and submm positions are
$0\fas08\pm0\fas07$ in right ascension and $0\fas33\pm0\fas05$ in declination.
The errors on the averaged offsets are estimated using bootstrapping,
where the 11 galaxies are resampled 1000 times.
We model the offsets between the sources as the systematic offset,
combined with one random offset for each source between the optical and submm position,
plus the error of the position measurement for the submm position.
We find that the offset between the submm and measured optical position can be modelled as a circular Gaussian
with a FWHM of $0.36^{+0.06}_{-0.09}$.
Again the errors are estimated based on bootstrapping,
where the 11 galaxies are resampled 1000 times.

\subsection{Robustness of the stacked data}
\label{sec:robust}
\begin{figure}
	\vbox to100mm{\vfil \includegraphics[width=85mm]{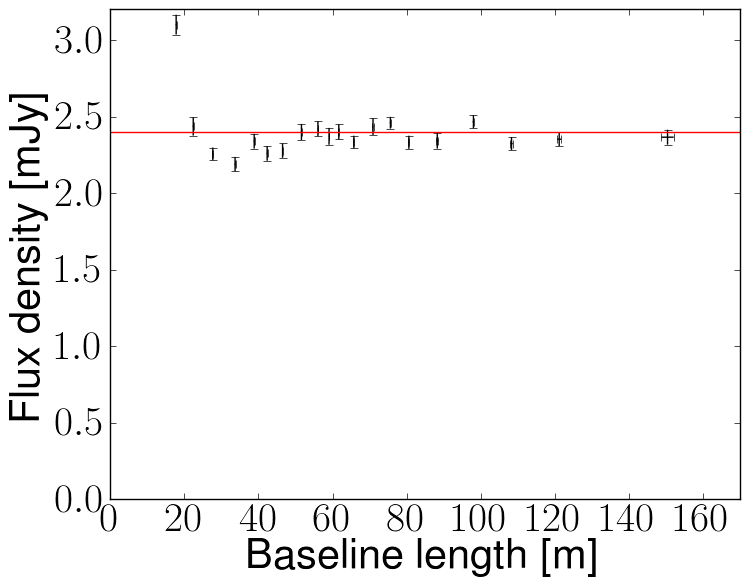}
	\caption{Stacked flux densities for a simulated dataset,
		produced by inserting point sources into the ALESS data.
		Flux densities averaged over 100 simulated datasets
		accurately estimate systematic biases.
		The noise is estimated as the standard deviation between the different simulations.
		The red line indicates the expected flux density for the stacked point sources.
		The shortest baseline is higher than the expected flux density due to contributions 
		from residuals of bright sources, see Lindroos et al. (2015) for more discussion of such effects.
	\label{fig:bias_estimate} }
	\vfil}
\end{figure}

To ensure robustness of our new results based on $uv$-stacking,
we perform several test on the stacked data and method.
By inserting and stacking point sources in the ALESS data,
using the method described in section \ref{sec:uncertainties},
we evaluate biases in the stacking result.
We find that the flux density agrees with the expected values,
except for the very shortest baselines,
where the flux density is approximately 20 per cent too high.
The results for the sBzK sample is shown in Fig. \ref{fig:bias_estimate},
however, the other samples show very similar structure in the $uv$-plane.
\cite{lindroos2015} found similar biases on the shortest baselines for simulated datasets.
In \cite{lindroos2015} this could be shown to be due to nearby bright sources which were not fully subtracted.
This is consistent with our data,
as the bright source subtraction is based on the clean models,
which may not fully subtract the sources.
Based on this we flag all baseline shorter than $18.2\,\mathrm{m}$ for the sBzK and DRG samples ($\sim$ 3 per cent of the data),
and all baselines shorter than $21\,$m for the \ksmp{} and ERO samples ($\sim$ 6 per cent of the data).

\begin{figure}
	\vbox to100mm{\vfil \includegraphics[width=85mm]{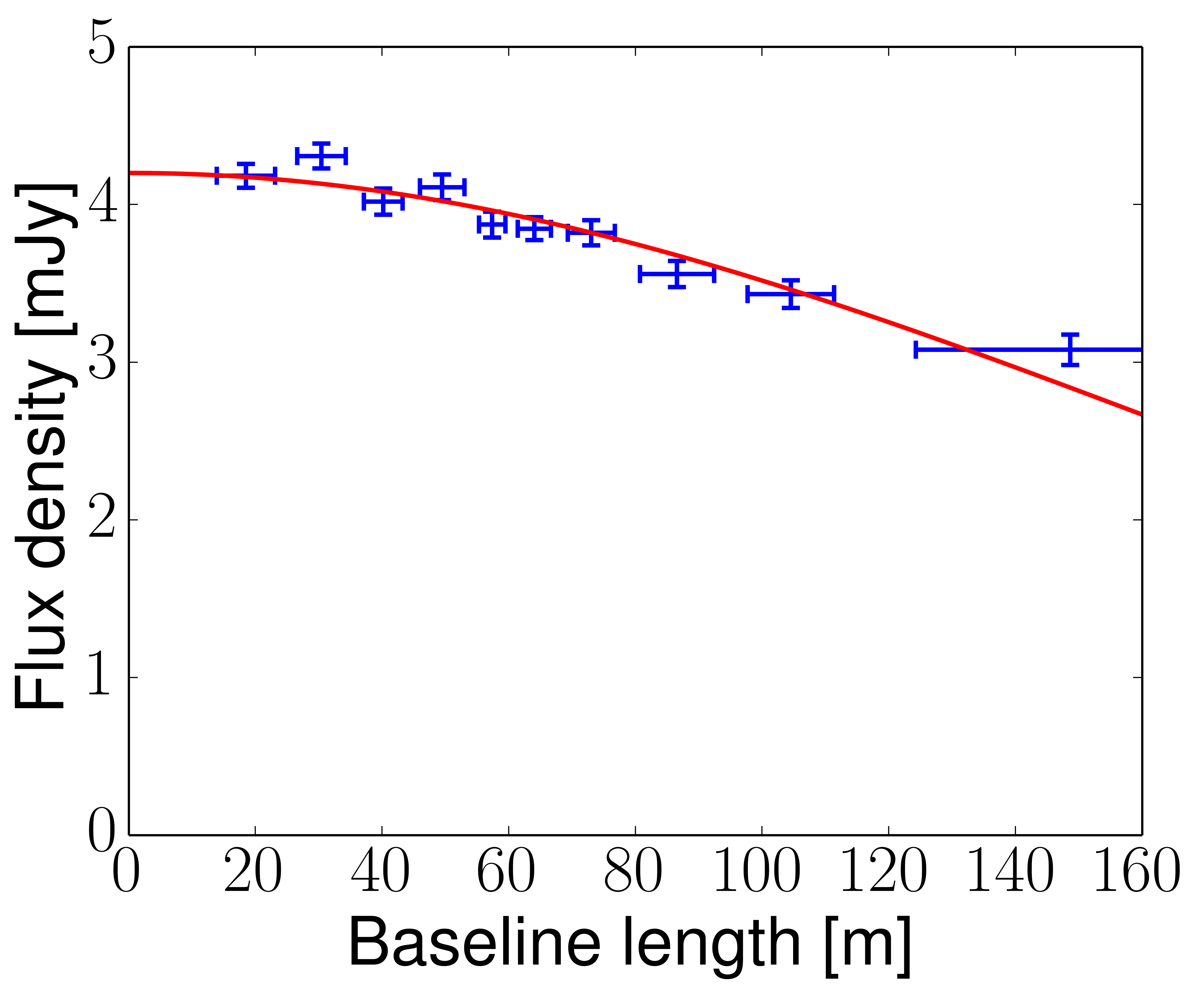}
		\caption{Flux densities for the stacked visibilities of the SMG sample.
			The visibilities are binned by baseline length.
			The red line indicates a Gaussian fit.
			The errors are estimated from the standard deviation of the real part of the visibilities within each bin.
			The horizontal error is estimated from the standard deviation of the $uv$-distance within each bin.
		\label{fig:smg_stacked}}
	\vfil}
\end{figure}

As an additional test we stack all SMGs in the data.
We use the sub-mm positions from the main catalogue from \cite{hodge2013} (99 sources).
This results in a peak SNR of $\sim60$.
The stacked data are well fitted by a Gaussian,
as shown in Fig. \ref{fig:smg_stacked},
with a flux density $4.2\pm0.14$ mJy and $0\fas4\pm0\fas06$.
This agrees well with \cite{simpson2015},
which found typical sizes (FWHM) of SMGs between 0\fas3 and 0\fas4.

\subsection{Flux densities and sizes of the stacked sources}
\begin{figure}
	\vbox to120mm{\vfil \includegraphics[width=85mm]{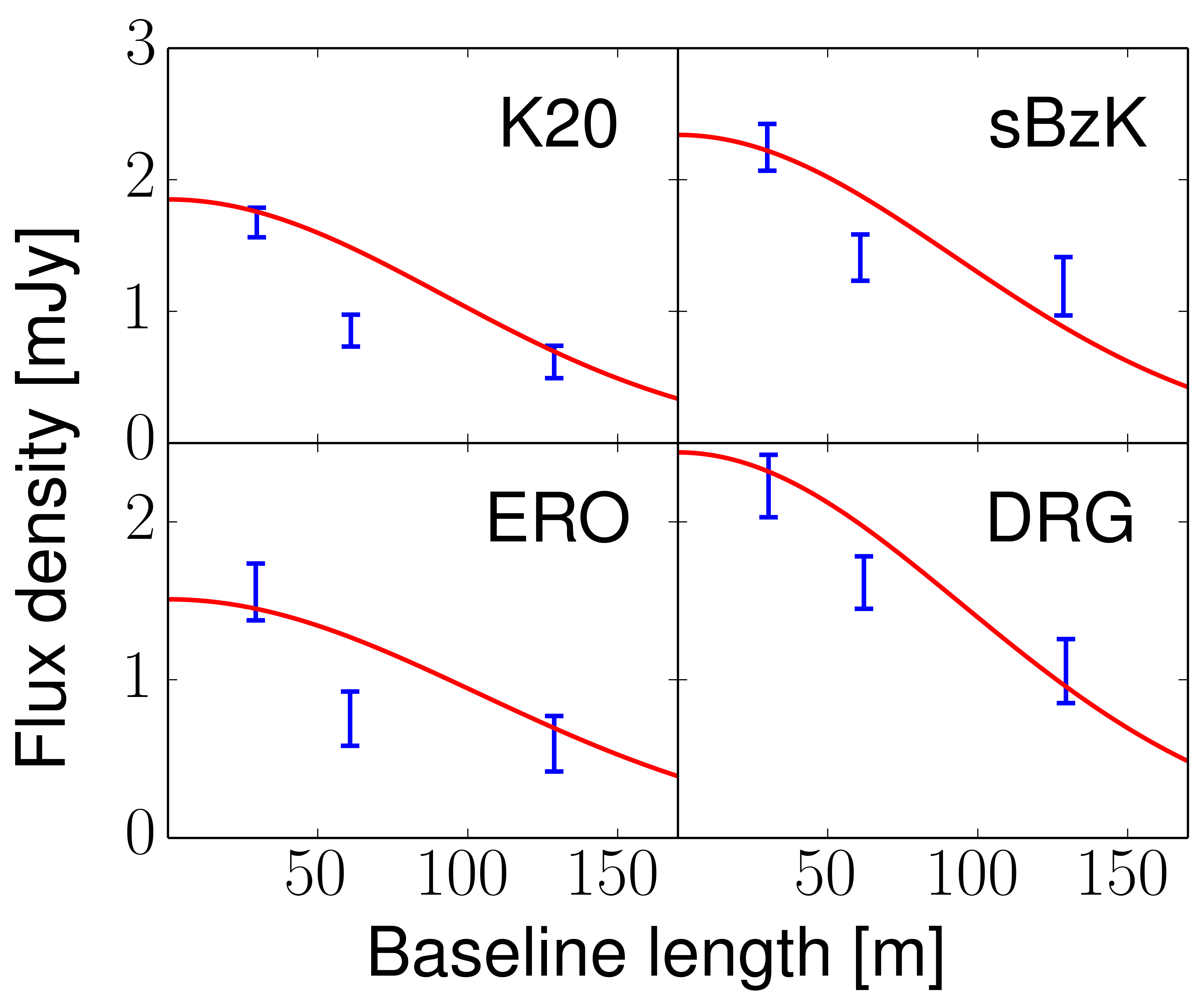}
	\caption{Stacked visibilities for each sample binned by baseline length.
	The errors are estimated from the standard deviation for the real part of the visibilities within each bin.
	The horizontal error is estimated from the standard deviation of the $uv$-distance within each bin.
	The lines show $uv$-models that are fitted to the full $uv$-data. 
	The blue dash-dotted line is a Gaussian, 
	the solid green line is a Gaussian plus a point source,
	and the black dashed line is a disk plus a point source.
	Note that no Gaussian model is visible for the DRG sample,
	as it is identical to the Gaussian + point source model for this sample.
	\label{fig:uvplot} }
	\vfil}
\end{figure}

Fig. \ref{fig:uvplot} shows flux density as a function of baseline length for each sample.
In the plot is shown the fit from the Gaussian model, with two free parameters:
the total flux density and the FWHM size.

The typical sources in all of our samples are found to be extended,
with stacked sizes between $0\fas65$ and $0\fas73$ (see Table \ref{tab:stack}).
The measured stacked sizes are broadened by random offsets between the measured $K$-band positions and submm positions.
Accounting for this effect,
we find deconvolved sizes for our samples between $0\fas54$ and $0\fas64$.
The uncertainties are estimated by using the bootstrap and Monte Carlo methods described in section \ref{sec:uncertainties}.
The bootstrapping errors are larger as they account for variance within the selected sample as well as observational uncertainties,
while the Monte Carlo only accounts for observational uncertainties.
For the deconvolved sizes, the reported errors are the combination of the Monte-Carlo errors and the errors on random offset measurements,
assuming that these two errors are independent.

Roughly half the galaxies in our samples are detected in the \emph{HST} $z$-band observations from GOODS-S and GEMS.
By fitting a S\'ersic distribution to these sources we can estimate the sizes at $z$-band wavelength.
We find a median size of 0\fas46 for the \ksmp{} sample and 0\fas52 for the other samples.
The median S\'erscic index $n$ is around 1.33 for each sample,
although slightly lower for the sBzK sample at 0.94.

Compared to the results from \cite{decarli2014},
we find flux densities which are 20 to 40 per cent higher.
This is expected as the image stacking method in \cite{decarli2014} uses the peak flux density in the stacked stamp,
which assumes that the sources are unresolved at the image resolution of 1\fas6.
When fitting a point source model to our $uv$-stacked data,
the measured flux densities deviate from the \cite{decarli2014} measurements by less than a few per cent.

\begin{table*}
	\centering
	\begin{minipage}{140mm}
	\caption{ \label{tab:stack}Flux density estimates with $uv$-stacking.
	The flux density in $uv$-stacking is estimated using two different methods.
	Method one (model): the flux density is estimated as the best fit Gaussian model.
	Method two (point source): the flux density is estimated as the weighted average of all unflagged visibilities.
	These two estimates would coincide for point sources.
	We also present the fitted size of the Gaussian model,
	as well as fitted size deconvolved from the random offsets between opical and sub-mm positions.
	For comparison the table also shows the image stacking results from Decarli et al. (2014).
	The errors are estimated by stacking fake sources introduced into the data.}

	\centering
	\begin{tabular}{c c C{1.8cm} C{1.5cm} C{1.8cm} C{1.8cm} C{2.4cm}}
		\\
		\hline
		\hline
                      &       & \multicolumn{4}{c}{$uv$-stacking }                                        & Image stacking     \\
		              &       & Gaussian      &                     &                     & Point source  &                    \\
		Sample        & N.gal & flux density  & Size                & Deconvolved         & flux density  & Peak flux density \\
		              &       & [mJy]         &                     & size                & [mJy]         & [mJy]             \\
		\hline                                                                                                            
		\ksmp         & 52    & $1.85\pm0.30$ & $0\fas73\pm0\fas14$ & $0\fas64\pm0\fas16$ & $1.14\pm0.07$                   & $1.16\pm0.09$            \\
		sBzK          & 22    & $2.34\pm0.32$ & $0\fas73\pm0\fas15$ & $0\fas64\pm0\fas17$ & $1.83\pm0.10$                   & $1.89\pm0.15$            \\
		ERO           & 25    & $1.51\pm0.22$ & $0\fas65\pm0\fas17$ & $0\fas54\pm0\fas19$ & $1.12\pm0.10$                   & $1.15\pm0.09$            \\
		DRG           & 19    & $2.44\pm0.28$ & $0\fas71\pm0\fas14$ & $0\fas61\pm0\fas16$ & $1.89\pm0.11$                   & $1.90\pm0.13$            \\
		\hline
	\end{tabular}
\end{minipage}
\end{table*}

\begin{table*}
	\centering
	\begin{minipage}{140mm}
		\caption{ \label{tab:stack_bootstrap} Distributions of stacked parameters as estimated from bootstrapping, resampling the galaxies within each sample 1000 times.
			These distributions include both errors from measurement uncertainties and variance within the samples.
			The presented range of 15.9 per cent to 84.1 per cent corresponds to the $\pm1\sigma$ range for a Gaussian distribution.
			The distributions are also presented as histograms in \ref{app:bootstrap}.
		}

	\centering
	\begin{tabular}{ccccccccccccc }
		\\
		\hline
		\hline
		Sample      & & \multicolumn{3}{c}{Gaussian flux [mJy]}& & \multicolumn{3}{c}{Size}   & & \multicolumn{3}{c}{Point source flux [mJy]} \\
					& & 15.9\% & 50\% & 84.1\%                 & & 15.9\%  & 50\%    & 84.1\% & & 15.9\% & 50\% & 84.1\%        \\
		\hline
		\ksmp       & & 1.33   & 1.90 & 2.58                   & & 0\fas63 & 0\fas94 & 1\fas38& & 0.95   & 1.25 & 1.61          \\
		sBzK        & & 1.62   & 2.38 & 3.14                   & & 0\fas54 & 0\fas74 & 0\fas91& & 1.31   & 1.86 & 2.33          \\
		ERO         & & 1.03   & 1.56 & 2.20                   & & 0\fas48 & 0\fas76 & 1\fas05& & 0.80   & 1.14 & 1.56          \\
		DRG         & & 1.81   & 2.43 & 3.16                   & & 0\fas54 & 0\fas72 & 0\fas85& & 1.49   & 1.91 & 2.32          \\
		\hline
	\end{tabular}
\end{minipage}
\end{table*}

\subsection{Simulations}
\begin{figure}
	\vbox to100mm{\vfil \includegraphics[width=85mm]{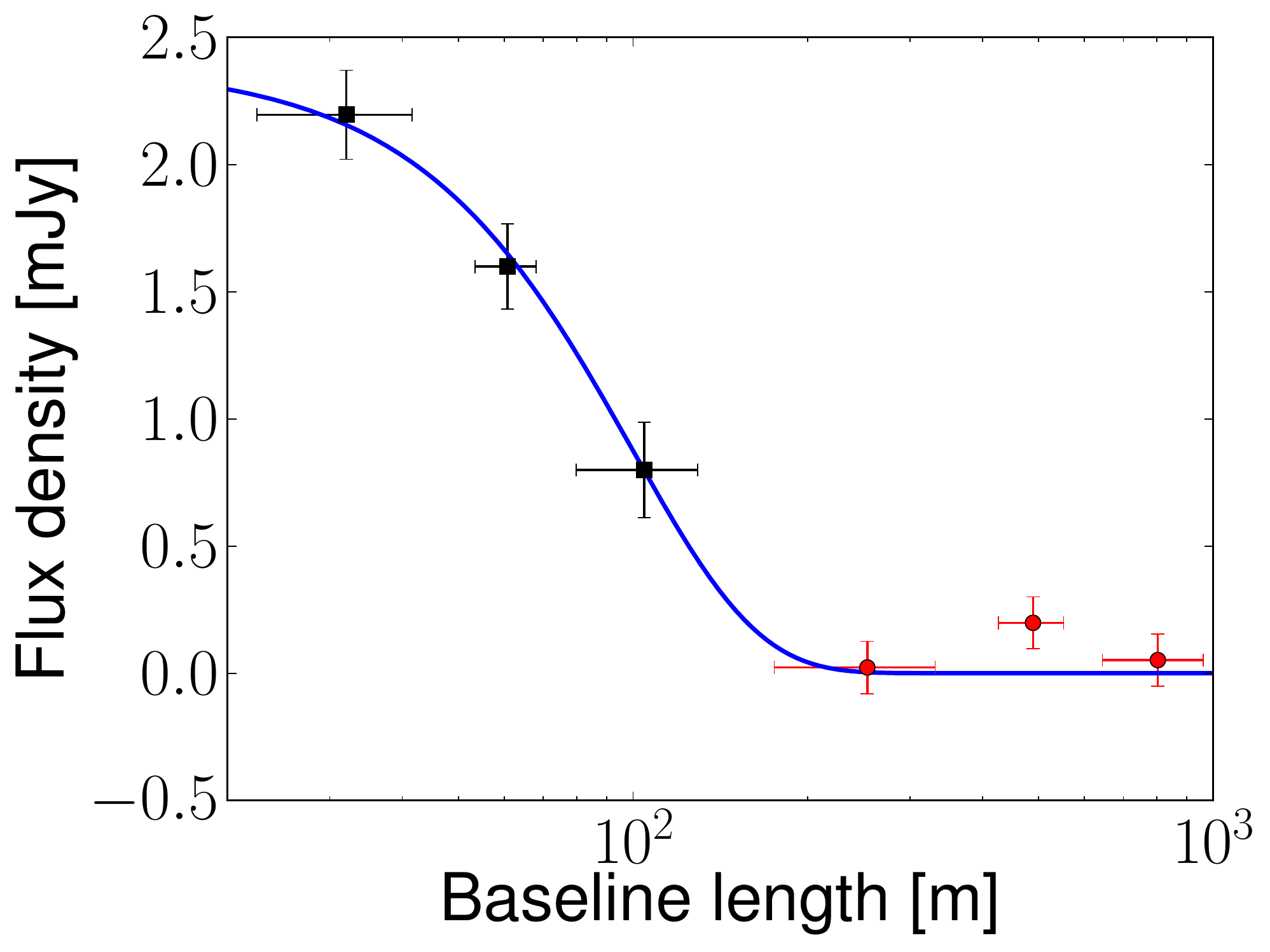}
	\caption{Stacked flux densities for simulated dataset.
		Each galaxy is simulated as a combination of three clumps,
		scattered within a radius of 5 kpc from the centre position for the galaxy.
		The errors are estimate from the standard deviations of the visibilities in each bin.
		The plot combines data from simulations with two different baseline configuration,
		The shorter baselines, marked with black squares,
		are simulated with the same $uv$-coverage as the ALESS observations.
		The longer baselines, marked with red circles,
		are simulated using an ALMA Cycle 3 configuration with baselines from 45 m to 1.4 km.
	\label{fig:clumps_lbi} }
	\vfil}
\end{figure}

To study the effect of substructure,
we perform a simulation in which the emission originates from kpc-scale clumps in the galaxies,
described in more detail in section \ref{sec:simulation}.
At baselines shorter than $\sim200$ m the stacked visibilities are well fitted by a Gaussian model,
as is shown in Fig. \ref{fig:clumps_lbi}.
The black squares indicate the ALESS baselines.
The simulation also include a set of longer baselines modelled on a 
intermediate length baseline configuration from ALMA Cycle 3,
with baselines from 45 m to 1400 m,
shown in Fig. \ref{fig:clumps_lbi} as red circles.
The Gaussian model recovers an average flux density for the stacked sources of $2.3\pm0.2$ mJy,
compared to the input flux density for the simulation of 2.1 mJy per source.
The flux density is primarily recovered by using the ALESS baselines,
using the long baselines from the ALMA Cycle 3 configuration,
we measure an average flux density of only 90 \uJy{}.
When fitting to the data from both baseline configurations,
the size measured for the Gaussian is 0\fas96$\pm0\fas30$.
This agrees well with the distribution of the positions for the clumps,
which are spread in a disk with a diameter of 1\fas2.

\begin{figure}
	\vbox to110mm{\vfil \includegraphics[width=85mm]{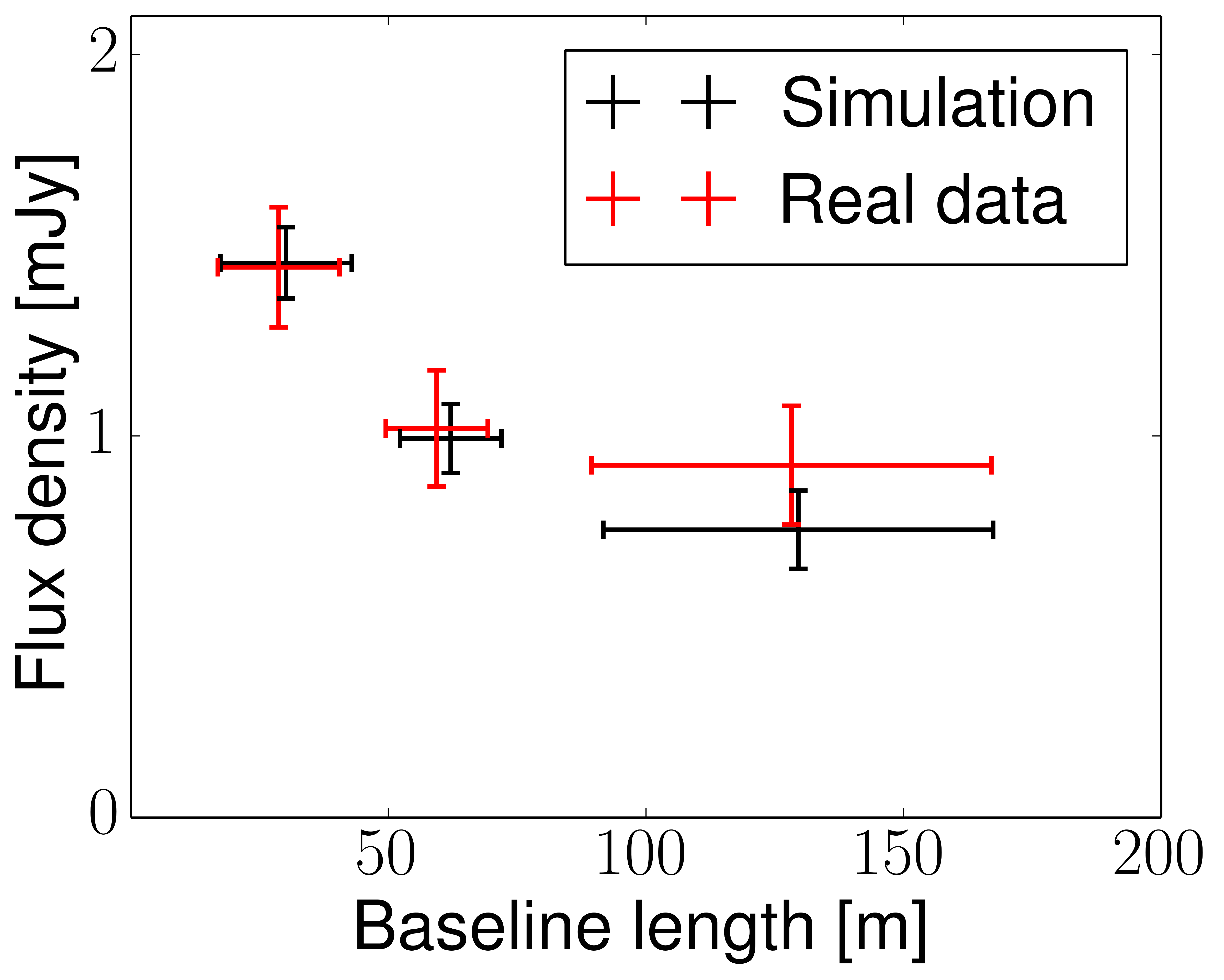}
	\caption{Simulation of stacked flux densities based on HST $z$-band emission maps shown in black,
		binned by baseline length.
		The errors are estimated from the standard deviation for the visibilities within each bin.
		For comparison the stacked flux densities of the $z$-band detected galaxies of our sample,
		using the same binning.
		Note that for the middle bin the simulated and real data are very close,
		and as such the simulated data point is hidden behind in the plot.
	\label{fig:hubble_sim} }
	\vfil}
\end{figure}

For the \emph{HST} $z$-band detected galaxies,
we measure and compare the HST sizes to our stacked ALMA sizes,
and find the values to be consistent with uncertainties for all samples.
However, for those sources with a strong detection we can perform a more in-depth comparison.
We select all sources from the \ksmp{} sample with peak SNR $>5$ in $z$-band,
a total of 32 sources.
Stacking these sources in the ALESS data we measure an average size of 0\fas77$\pm$0.15,
which compares well to the median effective radius at $z$-band (0\fas46).
For a more detailed comparison we perform a simulation based on the $z$-band morphology,
described in detail in section \ref{sec:hstsim}.
Fig. \ref{fig:hubble_sim} show the results of this simulation compared to the actual stacked ALESS data.
The simulated data and the actual stacked ALESS data show very similar scaling in the $uv$-plane,
indicating that the $z$-band and the sub-mm emission trace a similar radial morphology.

\subsection{Star-formation rates}
\begin{figure}
	\vbox to100mm{\vfil \includegraphics[width=85mm]{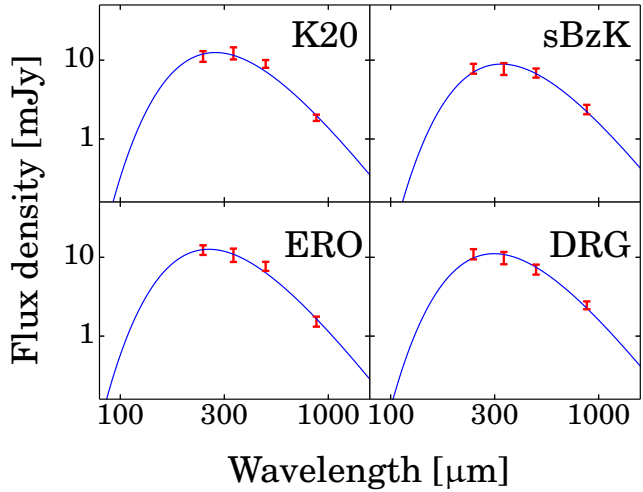}
		\caption{Stacked flux densities for the samples and fitted dust-emission SEDs. 
				 Combines the three wavelengths from the \emph{Herschel}/SPIRE with our new ALMA estimates.
			     The parameters of the fitted models can be found in Table \ref{tab:sfr}.
	\label{fig:dust_sed} }
	\vfil}
\end{figure}

\cite{decarli2014} stacked each of the four samples in the three \emph{Herschel} SPIRE bands.
Using data from the \emph{Herschel} Multi-tiered Extragalactic Survey \citep{oliver2012}.
We combine these values with our stacked ALESS flux densities to better constrain the dust spectral energy distributions (SED) of our samples.
The dust emission is modelled as a modified black body:
$S_\nu \propto \nu^\beta B_\nu(T)$ 
where $S_\nu$ is the dust SED,
$B_\nu(T)$ is the Planck function,
$T$ is the dust temperature (typically $T\approx$12-60 K),
and $\beta$ describes the effect of dust opacity (typically $\beta \approx 1.4-2$) \citep[e.g.][]{kelly2012}.
The total IR luminosity (\lir) is calculated between 8 \micron{} and 1000 \micron{} \citep[e.g.][]{sanders2003}.
The dust emission is fitted using a $\chi^2$ minimization,
with two free parameters, $T$ and \lir.
The value of $\beta$ is fixed to 1.6.
Each data point is weighted by $\sigma^{-2}$.
Data and fitted SEDs are shown in Fig. \ref{fig:dust_sed}, and results are summarised in Table \ref{tab:sfr}.

\begin{figure}
	\vbox to100mm{\vfil \includegraphics[width=85mm]{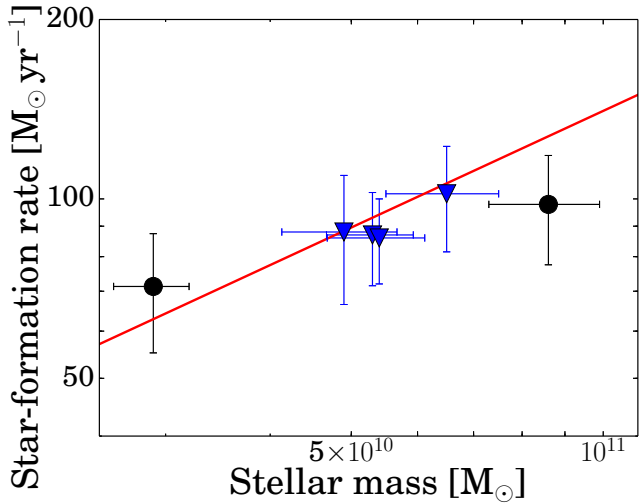}
		\caption{Average star-formation rate and stellar mass for the each sample shown as blue triangles (see Table \ref{tab:sfr}).
			The sBzK sample is also split into two sub-sample based on stellar mass,
            shown as black circles.
			The red line indicates the best-fit ``main sequence'' for star-forming galaxies at $z \sim 2$, 
			using the Tacconi et al. (2013) parametrization.
	\label{fig:SFR_vs_Ms} }
	\vfil}
\end{figure}

The SFRs are calculated from $\lir$ assuming a \cite{chabrier2003} initial mass function \citep{genzel2010}
\begin{equation}
	\mathrm{SFR} = 1.3\times 10^{-10}\,\msunyr \frac{\lir}{\lsun}.
\end{equation}
We find that the SFRs are similar for all samples at $\sim100$ \msunyr,
with the DRG sample showing a $\sim$20 per cent larger star-formation rate compared to the other samples.
In Fig. \ref{fig:SFR_vs_Ms} we show SFR as a function of stellar mass for each sample.
The measured values fall close to the best-fit ``main sequence'' for star-forming galaxies at similar redshifts,
(e.g. the \cite{tacconi2013} parametrization for comparison).
We also split the sBzK sample into two subsets based on stellar mass,
estimating the flux density of the stacked data with a Gaussian.
The star-formation rate is calculated using the same dust temperature as for the full sBzK sample.

\begin{table}
	\centering
	\caption{Infrared luminosity and SFR estimates for the stacked samples,
		using a combination of \emph{Herschel} and the new stacked ALMA results.
		We also show the average stellar mass for each sample.
		The errors are estimated from $\chi^2$ when varying both $T$ and $L_\mathrm{FIR}$ simultaneously.
	}
	\begin{tabular}{C{1.5cm} c c c c}
		\\
		\hline
		\hline
		Sample           & $L_\mathrm{FIR}$ & $T_\mathrm{dust}$  & SFR   & $M_*$  \\
		                 & [$10^{11}\mathrm{L}_\odot$] & [K] & [\msunyr] & [\msun] \\
		\hline
		\ksmp{}          & $6.9\pm1.4$   & $28\pm2$          & $90\pm18 $ & $5.3\times10^{10}$ \\
		sBzK             & $6.7\pm1.1$   & $27\pm2$          & $86\pm14 $ & $5.4\times10^{10}$ \\
		ERO              & $6.8\pm1.6$   & $30\pm3$          & $88\pm22 $ & $4.9\times10^{10}$ \\
		DRG              & $7.8\pm1.6$   & $28\pm2$          & $102\pm20$ & $6.5\times10^{10}$ \\
		\hline
		sBzK (high~mass) & $5.5\pm1.5$   & $27\pm2$          & $71\pm20 $ & $2.9\times10^{10}$ \\
		sBzK (low~mass)  & $7.6\pm1.5$   & $27\pm2$          & $98\pm20$  & $8.6\times10^{10}$ \\
		\hline
	\end{tabular}
	\label{tab:sfr}
\end{table}

\section{DISCUSSION}
\subsection{Extended emission}
Our stacked results show that the stacked sources have extended emission with typical sizes $\sim$0\fas7.
Assuming that the target sources are compact or unresolved,
as was done in \cite{decarli2014},
the flux density is systematically underestimated.
For the samples in this study with between 30 and 40 per cent. 
For the SMGs, where we measure the stacked size to be 0\fas4,
this effect is smaller with the peak brightness only $\sim$ 8 per cent lower than the full flux density.
Using model-fitting in the $uv$-domain we can effectively recover the full flux density.
This does, however, rely on having access sufficient sensitivity on short baselines.
The ALESS data were observed in a very compact ALMA configuration,
with most baselines shorter than 100\,m, or 115 k$\lambda$.
This results in a naturally weighted beam size of $\sim1\fas4$,
i.e., the observations are sensitive to scales of 1\arcsec{}-2\arcsec.

The filtering of spatial scales is a well known effect within interferometry,
however, the results of this study show that the effect is especially pronounced for stacking.
For the mapping of individual galaxies,
most of the flux density will originate from smaller scales,
allowing it to be resolved with higher resolutions.
Only emission which is smooth over larger scales is filtered.
In the case of stacking, the averaging of multiple galaxies smooth out substructure.
As such,
having access to sufficiently short baselines is essential to measure the total flux density of the stacked sources.
Emission at larger scales, at sizes larger than approximately 2-3\arcsec,
would be similarly suppressed in the ALESS data.
However, \emph{HST} data at $z$-band set an upper limit for our samples at around 2\arcsec,
as the dust-emission is unlikely to extend much beyond the stellar region.

\subsection{Robustness of the measured sizes}
Our simulations show that with stacking,
we can efficiently estimate the total flux density and the radial distribution of the emission.
Using Gaussian models,
we find sizes around 0\fas7 for the samples
and errors of 0\fas14-0\fas17.
This means that all samples are extended at a greater than $3\sigma$ significance.
\cite{vidal2012} calculate the limitation of model fitting of detected sources in a interferometric data set and
find that the minimal size that can be measured is given by
\begin{equation}
	\Theta_\mathrm{min} = \beta\left(\frac{\lambda_c}{2}\right)^{\frac{1}{4}} \left(\frac{1}{S/N}\right)^{\frac{1}{2}} \times \Theta_\mathrm{beam}
\end{equation}
where $S/N$ is the SNR of the averaged visibilities,
$\beta$ is a parameter that depends on the array configuration (typically between 0.5 and 1.0),
$\Theta_\mathrm{beam}$ is the FWHM of the beam using natural weighting,
and $\lambda_c$ depends on the probability cut-off for false detection (3.84 for $2\sigma$).
Using this formula we find our size error to be consistent with a $\beta$ between 0.4 and 0.5.
This both indicates that the sizes of $0\fas7$ are very robust,
and also shows that model fitting of stacked sources has similar noise to individual sources with similar SNR.
For comparison we also stacked the SMGs in our data, and find an average size of $0\fas4\pm0\fas1$.
This is marginally larger than the median size measured by \cite{simpson2015} of 0\fas3.
\cite{ikarashi2015} also measured the sizes of a sample of SMGs,
and find a smaller median size for the SMGs of 0\fas2,
however, with a different redshift distribution compared to our sample.

There are two factor which contribute to the measured sizes for our stacked sources:
the size of the galaxies and the random offsets between the optical and sub-mm positions.
Based on the brightest 11 sources in the \ksmp{} sample, which have a peak SNR $> 5 \sigma$,
we find that the typical offsets are $0\fas36\pm0\fas08$.
If we deconvolve this from the measured sizes we find that the sizes the actual galaxies are $0\fas54 - 0\fas64$.

We also estimate the variance of the target samples using bootstrapping.
This indicate larger errors on our estimated parameters due to the sample sizes,
with size errors increasing to 0\fas20 - 0\fas35.
Larger samples of star-forming galaxies have been studied using \textit{HST},
e.g. \cite{vanderwel2014} measured the sizes of $\sim 20000$ star-forming galaxies at $z > 1$.
Based on this sample they find that the optical sizes follow a log-normal distribution.
Looking at the sBzK galaxies,
if we assume that the sub-mm sizes of our samples follow a similar distributions,
we would expect this to contribute 0\fas04 to error of our stacked size assuming we sample 22 random galaxies.
This effect is similar for the other samples, getting smaller the larger the sample is.
Looking at results from bootstrapping,
we find that the results are consistent for the sBzK and DRG samples.
For the \ksmp{} the bootstrap estimated error is larger than expected from the optical sizes of star-forming galaxies,
however, this sample is not selective to star-forming galaxies leading probably leading to a more heterogeneous sample.
For the flux densities of our stacked sample, the bootstrap errors are larger than the measurement errors.
This is consistent with the large variation seen for star-forming galaxies,
where the SFR can vary with more than an order of magnitude within a sample.
We note that this indicates the error on the SFRs measured for our samples are dominated by sample variance.
This would be true even if each galaxy was individually detected,
indicating the importance of large samples to accurately estimate the typical SFR for a population of galaxies.

\subsection{Morphology of the underlying galaxies}
Looking at the sizes of the galaxies with a detection in the \textit{HST} $z$-band data (peak SNR $> 3$),
we can estimate the size of the stellar component of the galaxies.
Using a Sersic distribution, 
we find an median effective radius ($r_e$) of 0\fas5 with a median $n$ of 1.33.
The sizes measured at sub-mm wavelengths for our stacked sources are based on a Gaussian profile in place of a Sersic profile.
For comparison we fit our stacked sources using a Sersic profile, with $n$ fixed to 1.33,
and find that the sizes are consistent within errors as long as effective radius is compared to half the FWHM.
The difference is smallest for the sBzK sample at 3 per cent,
and largest for the DRG sample at 8 per cent.
Based on this analysis we find the measured sizes at sub-mm and optical wavelengths consistent within statistical uncertainties.

Approximately 70 per cent of our \textit{HST} observations are from the GEMS survey.
The GEMS $z$-band observations are not as deep as the GOODS-S $z$-band observations.
As such is possible that we are missing low flux surface density emission,
and underestimating the size of these galaxies.
However, as this primarily affects half the sample,
the impact on the median value is not expected to be very large.

Another limitation of the $z$-band measurements is dust obscuration.
The measured submm continuum emission indicates that dust is abundant in all samples.
We can compare to the shallower HST $H$-band observations from GEMS and GOODS-S,
which are less affected by dust absorption.
However, only 16 galaxies are detected in $H$-band.
For these galaxies we measure a median size of 0\fas6, which agrees well with the sizes measured in $z$-band.

The size of $0\fas7$ corresponds to a physical size of 6 kpc at the average redshift of the sBzK sample.
For SMGs several measurements of the sizes at sub-mm wavelengths exist,
e.g., \cite{simpson2015} find a median size of 2.4$\pm$0.6 kpc for SMGs with a median redshift of 2.6,
\cite{ikarashi2015} find size a median size of 0.7$\pm$0.13 kpc for galaxies with redshifts 3-6,
and \cite{hodge2015} measure the size of bright SMGs to $\sim$2$\times$1kpc.
In contrast, all our samples are significantly larger, with typical sizes which are more than twice as large.
For studies which select galaxies based on near-infrared (e.g. DRG and sBzK),
size measurements of sub-mm emission are more rare.
\cite{daddi2010} measure the sizes of 4 sBzK galaxies using IRAM Plateau de Bure Interferometer observations of the CO(2-1) transition,
and find sizes from 6 to 11 kpc (using a Gaussian model).
The \cite{daddi2010} detections have lower SNR than our stacked detection,
and the resolution of the observation is lower making the size estimate somewhat uncertain.
However, we can conclude that the results are consistent.

\subsection{Star-formation rate}
\subsubsection{Star-formation rate surface density}
Focusing on the sBzK sample,
the total SFR is estimated to be 100 \msunyr,
over a size of 10 kpc,
or a SFR surface density (\sfd) of 1 \sfdu.
This value is consistent with other measurements of sBzK galaxies,
e.g., \cite{daddi2010b} which found values for 0.1 to 30 \sfdu.
Of this, 40 per cent originates in the centre.
This corresponds to $\sfd \approx 13\,\sfdu$ in the inner 1 kpc of the galaxies.
While this is higher than the corresponding value for the DRGs ($\sim$ 2 \sfdu),
it is a very small value compared to LIRGs at lower redshift.
E.g., in Arp 220 with a similar SFR \citep{anantharamaiah2000}, 
the majority of the star formation occurs inside 1 kpc of the centre \citep{scoville1997},
resulting in an average \sfd{} of approximately 70 \sfdu \citep{anantharamaiah2000}.
We can also compare this to SMGs,
e.g., \cite{hodge2015} measured $\sfd$ in the centre of a $z=4$ SMG to be $\sim$120 \sfdu,
which is similar to Arp 220, but much higher than our sBzK galaxies.


As noted, \sfd{} in the centre of the DRG sample is very low,
at 2 \sfdu{} it is only a factor 4 above the same value for the Milky Way 
\citep{robitaille2009}, despite a factor 100 difference in SFR.

\subsubsection{SFR as a function of stellar mass}
In \cite{decarli2014}, all samples were found to have an excess of star formation compared to the similar samples in other fields.
Our updated flux-density estimate are $\sim$30 - 40 per cent higher than those found by \cite{decarli2014}.
However, after fitting the SED of the dust emission,
the fitted dust temperatures are typically lower.
For the sBzK and DRG samples,
this results in SFRs which are consistent with the \cite{decarli2014} measurements within the uncertainties.
However, for the \ksmp{} and ERO sample the SFR drops with $\sim 50$ per cent compared to \cite{decarli2014}.
This results in the \ksmp{}, ERO and sBzK samples having very similar star-formation rates,
at $\sim 90$ \msunyr.

We also compare the measured star-formation rates to the stellar masses,
and find them to be consistent with \cite{tacconi2013} for star-forming galaxies at $z\sim2$.
We also split the sBzK sample, the sample with highest SNR, by stellar mass.
Both the low- and high-mass samples fall close to the best-fit ``main sequence'' using the \cite{tacconi2013} parametrization.
This indicates, that while these galaxies are typically more massive compared to other similar samples,
the star formation is driven by the same mechanics.

%
%



%
%

%

\section{CONCLUSIONS}
In this paper we use stacking to measure the average morphologies and sizes of samples of galaxies using ALMA.
We use a $uv$-stacking algorithm combined with model fitting in the $uv$-domain.
We select star-forming galaxies at $z\sim{}2$ using four different criteria:
$K_\mathrm{VEGA} < 20$, ERO, DRG, and sBzK.
The samples are stacked in the ALMA 344 GHz continuum observations from the ALESS survey.
We find that all samples are extended,
with FWHM sizes of $\sim 0\fas7\pm0\fas2$ estimated using a Gaussian model.
Accounting for random offsets between optical catalogue positions and submm positions in the data,
we find that the actual average sizes are somewhat smaller at $\sim 0\fas6\pm0\fas2$.

The $uv$-model fitting results in flux densities that are $\sim40$ per cent higher than if the sources are assumed to be point sources.
Furthermore, assuming that the dust emission measured at 344 GHz is primarily heated by star formation,
we find that the majority of the star formation is taking place outside the inner kpc of the galaxy.
We compare this to the stellar distribution in the same galaxies,
using \emph{HST} $z$-band data.
The median effective radius is measured to 0\fas6,
which agrees well with the submm sizes.
We also simulate an ALMA data set with the rescaled $z$-band maps as input model for each galaxy.
The distribution are found to agree well,
indicating no systematic difference in size or radial distributions between the stellar and star-forming component.


Using a Monte Carlo method to estimate the robustness of the result,
we find the measured sizes to be robust at $>3 \sigma$ for all samples.
The measured difference between the sBzK and DRG sample,
is larger than the uncertainties with a statistical significance of $2\sigma$.
We find that the measured accuracy of the sizes is comparable to the theoretical limits for individual sources
\citep[e.g.][]{vidal2014}.
As in all cases with stacking we do not measure the properties of the individual galaxies,
but the average properties of the samples,
and this smoothing effect can simplify the modelling of the stacked source.
However, it also increase the interferometric effect of filtering of large spatial scale,
making short spacings very important to recover the full flux density.

We can conclude that for the stacking of any sources that may be marginally extended,
using $uv$-stacking with model fitting can provide a flux-density estimate that is significantly more robust
and valuable additional information such as the typical sizes of the sources of the stacked sample.
This is also important for future facilities such as the Square Kilometer Array (SKA),
showing that having access to $uv$-data in stacking is invaluable.

\section{Acknowledgements}
LL thanks Robert Beswick for useful discussion,
Ivan Mart\'{i}-Vidal for helpful input on the model fitting,
and Ian Smail for useful discussion.
We thank an anonymous referee for helpful suggestions and useful comments.
This paper makes use of the following ALMA data: ADS/JAO.ALMA\#2011.1.00294.S.
ALMA is a partnership of ESO (representing its member states), NSF (USA), and NINS (Japan),
together with NRC (Canada) and NSC and ASIAA (Taiwan), in cooperation with the Republic of Chile.
The Joint ALMA Observatory is operated by ESO, AUI/NRAO, and NAOJ.
This publication also makes use of data acquired with European Southern Observatories VLT under program ID 183.A-0666.
KK acknowledges support from the Swedish Research Council, and the Knut and Alice Wallenberg Foundation.
JLW is supported by a European Union COFUND/Durham Junior Research Fellowship under EU grant agreement number 267209, and acknowledges additional support from STFC (ST/L00075X/1).
The Dark Cosmology Centre is supported by the Danish National Research Foundation.
AK acknowledges support by the Collaborative Research Council 956, sub-project A1, funded by the Deutsche Forschungsgemeinschaft (DFG).
KC acknowledges support from UK’s Science and Technology Facilities Council (STFC) grant ST/M001008/1.

\label{lastpage}

\appendix
\bibliography{lindroos2016a} 

\section{Fitted models}
\label{app:bootstrap}
In this appendix we present the distributions determined for the fitted sizes using bootstrapping on the stacking samples.
The method for the bootstrapping is described in \S\ref{sec:uncertainties},
and the plotted distribution indicate the probability of possible sizes for the population of each sample.
The bootstrapping method approximate errors from observational noise as well as sample variance.

\begin{figure}
	\vbox to100mm{\vfil \includegraphics[width=85mm]{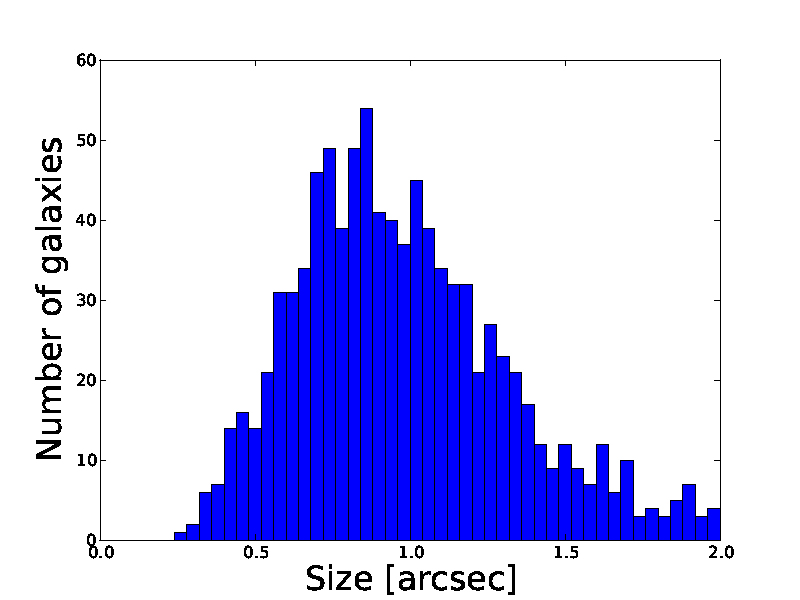}
		\caption{Distribution of stacked size for the \ksmp{} sample as estimated through bootstrapping.
		\label{fig:k20_size_ext}}
	\vfil}
\end{figure}

\begin{figure}
	\vbox to100mm{\vfil \includegraphics[width=85mm]{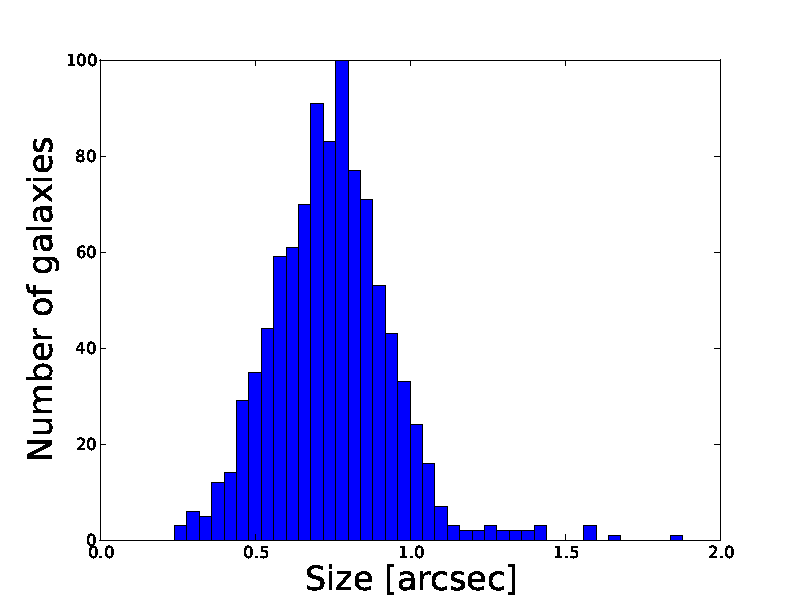}
		\caption{Distribution of stacked size for the sBzK sample as estimated through bootstrapping.
		\label{fig:sbzk_size_ext}}
	\vfil}
\end{figure}

\begin{figure}
	\vbox to100mm{\vfil \includegraphics[width=85mm]{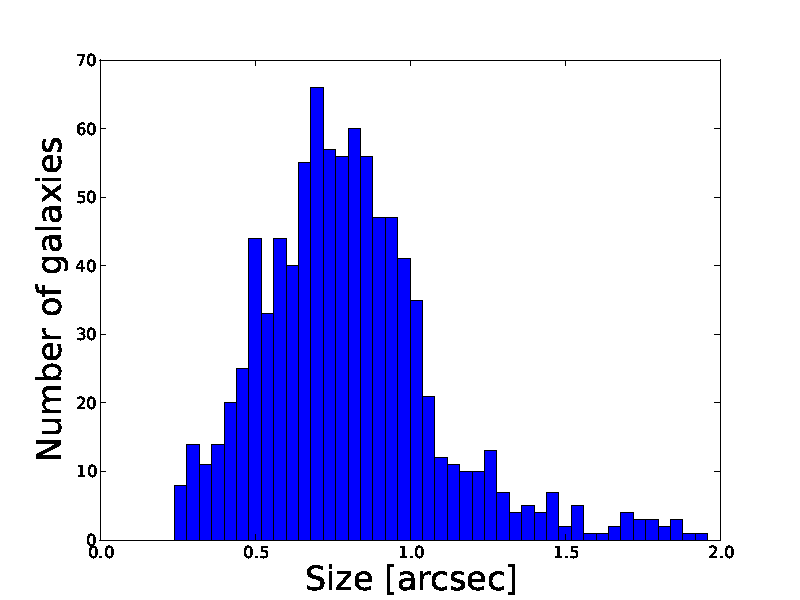}
		\caption{Distribution of stacked size for the ERO sample as estimated through bootstrapping.
		\label{fig:ero_size_ext}}
	\vfil}
\end{figure}

\begin{figure}
	\vbox to100mm{\vfil \includegraphics[width=85mm]{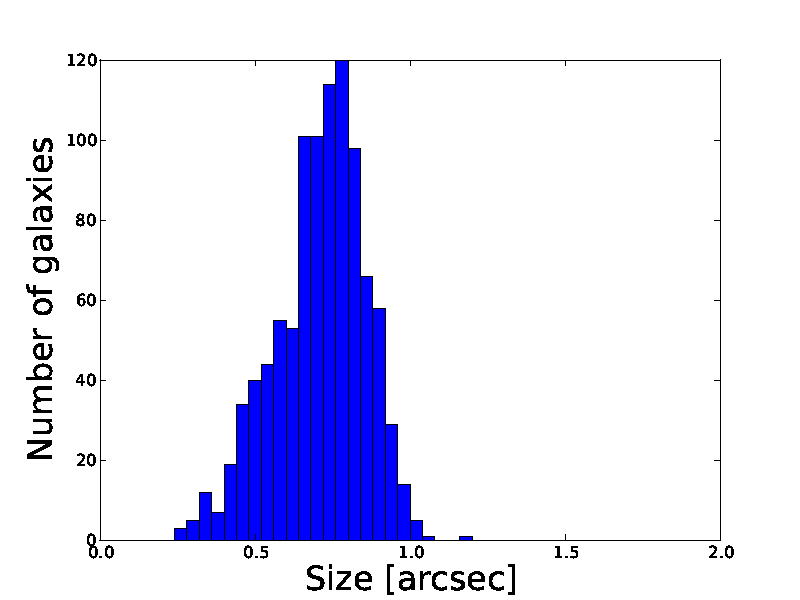}
		\caption{Distribution of stacked size for the DRG sample as estimated through bootstrapping.
		\label{fig:drg_size_ext}}
	\vfil}
\end{figure}

\end{document}